\documentclass[aps, prd, preprint
, nofootinbib, 
]{revtex4-1}

\usepackage{amsmath}
\usepackage{amsfonts}
\usepackage{bm}
\usepackage{color}
\usepackage{amssymb}
\usepackage{latexsym}
\usepackage{cleveref}
\usepackage{graphicx}
\usepackage{comment}
\usepackage{empheq}

\def\be{\begin{equation}}
\def\ee{\end{equation}}
\def\ben{\begin{eqnarray}}
\def\een{\end{eqnarray}}

\begin{document}

\preprint{KEK-QUP-2022-0004, KEK-TH-2467, KEK-Cosmo-0301}
\title{Superluminal propagation from IR physics}
\author{Asuka Ito}
\email[]{asuka.ito@kek.jp}
\affiliation{International Center for Quantum-field Measurement Systems for Studies of the Universe and Particles (QUP), KEK, Tsukuba 305-0801, Japan}
\affiliation{Theory Center, Institute of Particle and Nuclear Studies, KEK, Tsukuba 305-0801, Japan}

\author{Teruaki Suyama}
\email[]{suyama.t.ab@m.titech.ac.jp}
\affiliation{Department of Physics, Tokyo Institute of Technology, 2-12-1 Ookayama, Meguro-ku, Tokyo 152-8551, Japan}

\begin{abstract}
One may believe that front velocities of waves in a given theory coincide with the UV limit of phase velocities for any dispersion relations.
This implies that IR physics is irrelevant to the discussion of propagation speed of waves.
We first consider a theory that contains higher spatial derivatives in the wave equation and prove that front velocities coincide with the UV limit of phase velocities, at least, if parity is conserved.
However, we also show that front velocities do not coincide with the UV limit of phase velocities in general dispersion relations.
We explicitly give several examples in which front velocities are superluminal owing to an IR or intermediate energy scale property of dispersion relations even 
if the UV limit of phase velocities is luminal.
Our finding conveys the important caution that not only UV physics but also IR physics can be significant to superluminality.
\end{abstract}

\maketitle

\section{Introduction}
The superluminal propagation of waves in a given theory is usually considered to be an unacceptable nature; 
non-existence of the superluminal propagation can be a guiding principle in constructing a consistent theory%
\footnote{It is not obvious how superluminality is related to the violation of causality
when the Lorentz invariance is broken.
We simply stand at the position that we do not allow the existence of superluminality in a frame for the moment.}.
Indeed, superluminality can be akin to the illness of a theory~\cite{Aharonov:1969vu,CarrilloGonzalez:2022fwg}.
Although we often encounter apparent superluminality, that is, group velocities are faster than the speed of light, 
in various realistic situations, such as photons propagating in materials~\cite{5535097} or an axion dark matter background~\cite{Adshead:2020jqk}, and 
lensed gravitational waves~\cite{Suyama:2020lbf,Ezquiaga:2020spg},
they do not mean true superluminal propagation.
Actually, front velocities, which are more suited to
quantifying propagation speed, are inside a light cone in the above situations~\cite{5535097,Adshead:2020jqk,Suyama:2020lbf,Ezquiaga:2020spg}.

As one investigates superluminality with front velocities in a theory,
the relation 
\begin{equation}
  v_{f} = \lim_{k \to \infty} v_{p}(k) ,  \label{rela} 
\end{equation}
may be useful. Here, $k$ represents the wavenumber (=momentum), and 
$v_{f}$ and $v_{p}(k) \equiv \frac{\omega_k}{k}$ are the front velocity and the phase velocity of a wave
obeying the dispersion relation $\omega=\omega_k$, respectively.
So far, to our best knowledge, Eq.\,(\ref{rela}) has been proved only for limited cases where the dispersion relation has the following 
form:
\begin{equation}
  \omega^{2}_{k} = k^{2} + (\sigma) k^{\alpha}M \quad (\alpha = 0,1,2),  \label{pow}
\end{equation}
where $\sigma$ takes $+$ or $-$ and $M$ is a positive constant ($M<1$ for $\alpha = 2$).
The case of $(\sigma=+,\ \alpha=0)$ represents a standard dispersion relation for a massive particle. 
On the other hand, $(\sigma=-,\ \alpha=0)$ represents a particle with a tachyonic mass.
In this case, it was shown, by evaluating a Green function, that although the group velocity is superluminal, 
the front velocity coincides with the speed of light~\cite{Aharonov:1969vu}.
For photons propagating in an axion dark matter background, the dispersion relation with $(\sigma=\pm,\ \alpha=1)$ is realized for timescales that are short compared with the period of the axion oscillation.
Then apparently, the propagation seems superluminal because the group velocity is superluminal; however, one can see 
that the front velocity is luminal by solving a Green function~\cite{Adshead:2020jqk}.
Therefore, the above works~\cite{Aharonov:1969vu,Adshead:2020jqk} prove relation (\ref{rela}) 
when the dispersion relation is given by Eq.\,(\ref{pow}).
We note that Eq.\,(\ref{rela}) also holds for the $\alpha = 2$ case trivially, where the phase velocity 
in the UV limit coincides with that in the IR limit.
Furthermore, more general discussion about partial differential equations up to the second order shows that
Eq.\,(\ref{rela}) 
is valid in the case of Eq.\,(\ref{pow})~\cite{mandelshtamlectures,Shore:2007um}.

Eq.\,(\ref{rela}) indicates that the UV limit of a phase velocity determines the front velocity.
Actually, this feature plays an important role when we consider low energy effective field 
theories from the viewpoint of UV completion, as discussed in%
\footnote{It may be natural to consider a UV cutoff when one considers low energy effective field theories.
Then, mild superluminality (i.e., consistent with luminal propagation 
within the range of the effective field theory which has limiting resolution) is 
allowed~\cite{deRham:2021bll,Goon:2016une,Chen:2021bvg,deRham:2020zyh,CarrilloGonzalez:2022fwg} 
in principle.
Even for such a case, in general, one needs to investigate front velocities (though they now have the uncertainty corresponding to the cutoff) rather than group or phase velocities to evaluate the propagation speed, as discussed in~\cite{Suyama:2020lbf,Ezquiaga:2020spg}.}~\cite{Shore:2007um,deRham:2019ctd}.
Remarkably, Eq.\,(\ref{rela}) indicates that IR physics is irrelevant to the propagation speed.
One might think that this is counterintuitive since it means that
even the propagation speed of waves consisting of low-frequency modes
is specified by only the dispersion relation for the very high frequency modes rather than 
by the dispersion relation 
for the modes contained by the waves.
This motivates us to investigate whether Eq.\,(\ref{rela}) is true not only for the dispersion relation 
of the form given by Eq.~(\ref{pow}) but also for other functional forms.
As we will demonstrate, Eq.\,(\ref{rela}) does not always hold true.

In this paper, we first show that relation (\ref{rela}) is true even when 
$\alpha = 4, 6, 8, \dots$ in Eq.\,(\ref{pow}).
This proof extends the applicable range of relation (\ref{rela}) and supports the discussion about 
low energy effective field theories from the viewpoint of UV completion~\cite{Shore:2007um,deRham:2019ctd}.
Furthermore, we also provide explicit examples that violate Eq.\,(\ref{rela}).
Our analysis reveals that not only UV physics but also IR or intermediate-scale physics can be relevant 
to the discussion of propagation speed in general.

\section{Infinite front velocities from UV physics}  \label{UV}
To simplify the computation without losing the essential point,
throughout this paper, we assume that the dispersion relation is obeyed by a free scalar field $\phi$.
Given that any wave is given by a superposition of plane waves, it is sufficient to restrict the 
dimension of space to one for the purpose of determining the front velocity.
Thus, practically, our spacetime is two dimensions whose coordinates are $(t,x)$.
With this premise, the equation of motion for $\phi$ is given by
\be
\left( -\frac{\partial^2}{\partial t^2} +\omega^2 (-\partial_x^2) \right)
\phi (t,x)=0.  \label{free-eq}
\ee
Here, $\omega^2 (X)$, which defines the dispersion relation, is a function of $X$.
We assume that the wave equation respects parity. In other words,
the wave equation contains only an even number of the spatial derivatives. 
Thus, $\omega^2$ is a function of $-\partial_x^2$.

In this section, we will show that Eq.\,(\ref{rela}) with dispersion relation (\ref{pow}) 
is valid, even when $\alpha$ is larger than 2, by explicitly evaluating the (retarded) Green function.
In this case, the wave equation contains higher-order derivative terms in space coordinate $x$.
To construct the Green function, let us add the local source term in the 
right-hand side of the wave equation (\ref{free-eq}):
\be
\left( -\frac{\partial^2}{\partial t^2} +\omega^2 (-\partial_x^2) \right)
\phi (t,x)=-\delta (t) \delta (x).  \label{eq}
\ee
To solve this equation, let us Fourier-transform $\phi$ as
\be
\phi (t,x)=\int \frac{dk}{2\pi} e^{ikx}\phi_k (t).
\ee
Then, the wave equation is transformed as
\be
\left( \frac{d^2}{dt^2}+\omega_k^2 \right) \phi_k (t)=\delta (t),
\ee
where $\omega_k^2 \equiv \omega^2 (k^2)$.
It can be verified that the inhomogeneous solution that vanishes for $t<0$
is given by
\be
\phi_k (t)=\theta (t)\frac{\sin (\omega_k t)}{\omega_k},
\ee
where $\theta (t)$ is the Heaviside function.
It should be understood that $\omega_k $ is the positive 
solution of $\omega_k^2 \equiv \omega^2 (k^2)$, i.e., 
$\omega_k =\sqrt{\omega^2(k^2)}$.
Thus, the desired solution is given by
\be
\phi (t,x)=\theta (t) \int \frac{dk}{2\pi} e^{ikx}\frac{\sin (\omega_k t)}{\omega_k}.
\ee
For $t>0$, which is the domain of our interest, 
the time derivative of $\phi$ becomes
\be
{\dot \phi} (t,x)= \int \frac{dk}{2\pi} e^{ikx}\cos (\omega_k t)
=\frac{1}{2} \int \frac{dk}{2\pi} e^{ikx-i\omega_k t}
+\frac{1}{2} \int \frac{dk}{2\pi} e^{ikx+i\omega_k t}.   \label{time}
\ee
Using the fact that $\omega_k$ is an even function of $k$, we have
\be
\frac{1}{2} \int \frac{dk}{2\pi} e^{ikx+i\omega_k t}=
\frac{1}{2} \int \frac{dk}{2\pi} e^{-ikx+i\omega_k t}
=\left( \frac{1}{2} \int \frac{dk}{2\pi} e^{ikx-i\omega_k t} \right)^*.
\ee
Thus, ${\dot \phi}$ becomes
\be
\label{s2:formal-phi}
{\dot \phi} (t,x)
={\rm Re} \int \frac{dk}{2\pi} e^{ikx-i\omega_k t}.
\ee
Until $t=0$, $\phi=0$ at any location. 
After $t=0$, the value of $\phi$ may change from the initial value 
because of the source term
$-\delta (t) \delta (x)$.
Since the change of $\phi$ at a fixed point $x$ is caused by  
the nonvanishing of ${\dot \phi}$, we can say that 
the wave front at $t \ (>0)$ is the place where ${\dot \phi}$ has just
become nonzero for the first time at $t$. Then, the front velocity is obtained simply 
by dividing the distance between that place and the origin by $t$.

In what follows, we will show that the front velocity is infinite if
$\omega_k$ grows faster than $k$ as $k \to \infty$.
To this end, we first assume that $\omega_k$ approaches a power-law form as 
$k \to \infty$:
\be
\label{asymptotic}
\lim_{k\to \infty} \frac{\omega_k}{k^n} =\frac{\beta}{n},
\ee
where $\beta$ is some constant and $n>1$.

If we take $t \ (>0)$ to be sufficiently small, then
the phase $\omega_k t$ in Eq.~(\ref{s2:formal-phi}) remains negligibly small until $k$ becomes
extremely large, at which point we may safely use the asymptotic form (\ref{asymptotic})
for $\omega_k$.
Thus,
\be
I \equiv \int \frac{dk}{2\pi} e^{ikx-i\omega_k t}
=\int \frac{dk}{2\pi} e^{ikx-i \frac{\beta}{n}k^n t}  \label{III}
\ee
holds true for sufficiently small $t$.
For technical convenience, let us write $t$ as $t=\epsilon^{n-1}$.
Since $n>1$, $\epsilon$ is a very small number as well.
Plugging $t=\epsilon^{n-1}$ into Eq.\,(\ref{III}) and rescaling the integration variable $k$, we obtain
\be
I=\frac{1}{\epsilon}
\int \frac{d\kappa}{2\pi} \exp \left( \frac{i}{\epsilon} f(\kappa) \right),~~~~~~~~~
f(\kappa) \equiv x\kappa -\frac{\beta}{n} \kappa^n.
\ee
Now, when $\epsilon$ is sufficiently small, the integrand is a highly
oscillating function except at the point $\kappa_0$ where the phase $f(\kappa)$ becomes stationary, i.e., $f'(\kappa_0)=0$. 
Then, $I$ is dominated by the integration around the stationary point.
On the basis of this observation, let us Taylor-expand $f(\kappa)$ around $\kappa=\kappa_0$ and truncate it at the leading order:
\be
f(\kappa)=f(\kappa_0)+\frac{f''(\kappa_0)}{2} {(\kappa-\kappa_0)}^2.  \label{statio}
\ee
Using Eq.\,(\ref{statio}), we can perform the integration 
analytically
and we end up with
\be
I=\frac{1}{\sqrt{2\pi \epsilon |f''(\kappa_0)|}} \exp \left(
\frac{i}{\epsilon} f(\kappa_0)-i \frac{\pi}{4} \right).
\ee
Using the explicit form of $f(\kappa)$, we find
\be
f(\kappa_0)=\left( 1-\frac{1}{n} \right)
{\left( \frac{x^n}{\beta} \right)}^{\frac{1}{n-1}},~~~~~
f''(\kappa_0)=-(n-1) {\left( \beta x^{n-2} \right)}^{\frac{1}{n-1}}.
\ee
Obviously, the real part of $I$ does not vanish.
This means that ${\dot \phi}$ is nonzero at any point $x$ infinitesimally 
after $t=0$.
Thus, the front velocity is infinite.
This result is consistent with Eq.\,(\ref{rela}) because now the phase velocity is also infinite in the
UV limit as expressed by Eq.\,(\ref{asymptotic}).
Therefore, relation (\ref{rela}) is
true even for $\alpha = 4, 6, 8, \dots$ in Eq.\,(\ref{pow}).

\section{Superluminal propagation from IR or intermediate-scale physics}   \label{IR}
In the previous section, we showed that front velocities are infinite if phase velocities increase infinitely in the UV limit by 
considering the UV modification of the dispersion relation as Eq.\,(\ref{free-eq}).
This can be regarded as a generalization of relation (\ref{rela}) with dispersion relation (\ref{pow}).
One then may believe that front velocities only depend on UV physics, and thus, IR or intermediate-scale physics is irrelevant to propagation speed.
However, this naive expectation is not true in general.
Below, we will show several examples that have infinite front velocities, even though their dispersion relations are modified only in 
IR or intermediate scales.

\subsection{Discontinuous dispersion relation}
\begin{figure}[ht]  
\centering
\includegraphics[width=6cm]{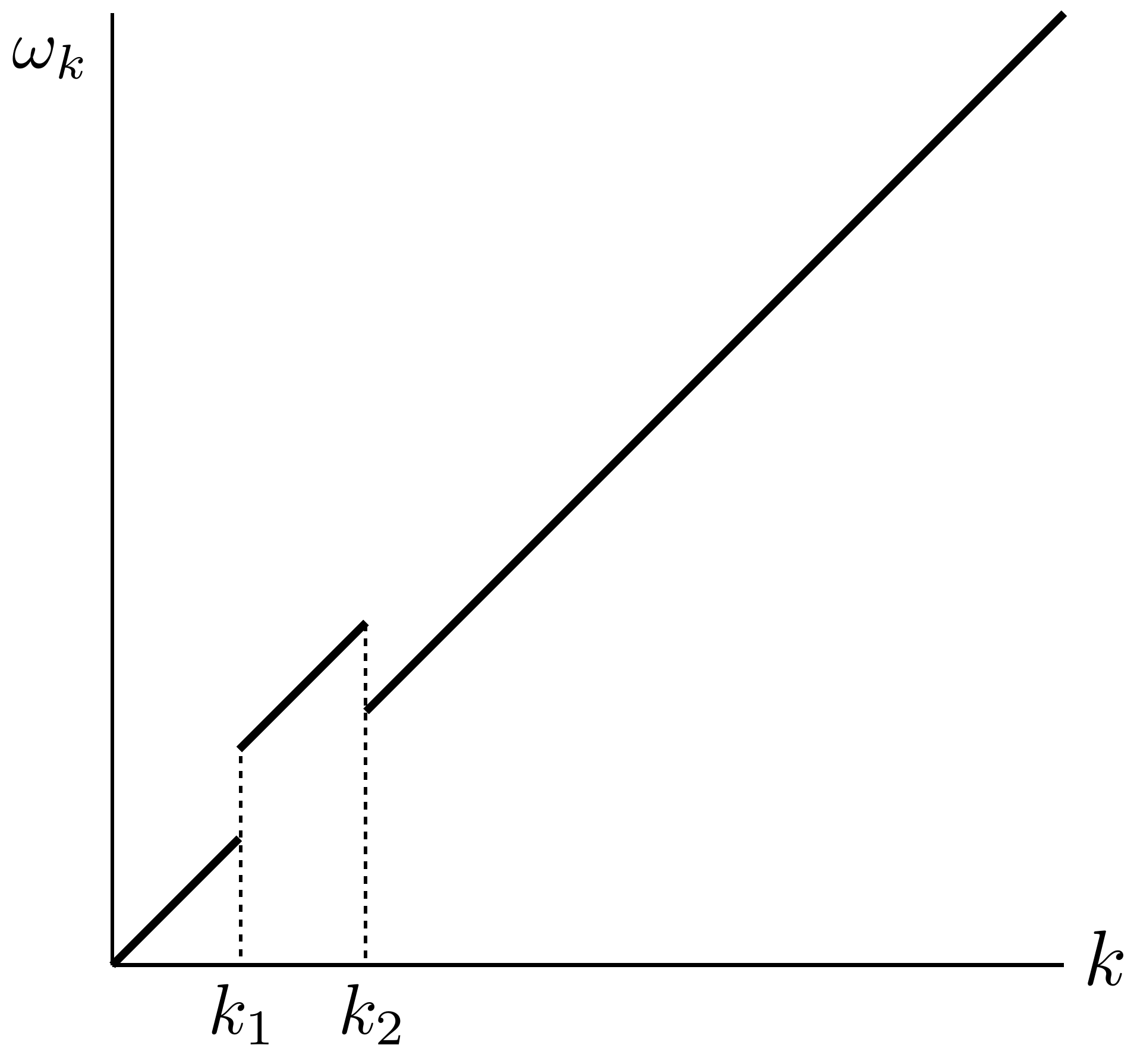}
\caption{Depiction of dispersion relation (\ref{dis0}).
} \label{fig_tobi}
\end{figure}

Let us first consider a dispersion relation that has discontinuities at momenta $k_{1}$ and $k_{2}$, as depicted
in Fig.\,\ref{fig_tobi}.
Hereafter, we will assume that parity is conserved, i.e., $\omega_{-k} = \omega_{k}$, as in Sec.\,\ref{UV}.
For $k>0$, the dispersion relation is given by
\begin{equation}
 \omega_{k} =
\begin{dcases*}
    k & for \  $ k < k_{1}, \quad k_{2} < k$ \ , \\
    k + C   & for \  $k_{1} < k < k_{2}$ \ ,
\end{dcases*}  \label{dis0}
\end{equation}
where $C$ is a constant.
This dispersion relation is highly artificial.
Nevertheless, we consider this example first because (i) simple analytic calculation is possible
and (ii) it nicely represents that modification of the dispersion relation only in the IR regime still leads to infinite front velocity.
To investigate front velocity, as was done in Sec.\,\ref{UV},
we consider the point source with the wave equation that has the above dispersion relation.
Then we see that the time derivative of the solution is given by Eq.\,(\ref{time}) by repeating the previous discussion in Sec.\,\ref{UV}.
From Eqs.\,(\ref{time}) and (\ref{dis0}), we obtain
\begin{eqnarray}
  {\dot \phi} (t,x) 
               =  \int^{\infty}_{-\infty} \frac{dk}{2\pi} e^{ikx} \cos (kt)  
                      - \int^{k_{2}}_{k_{1}} \frac{dk}{\pi} \cos(kx) 
                        \Big[ \cos(kt) - \cos\big[(k+C)t \big] \Big] .  \label{inttobi}
\end{eqnarray}
The first term yields delta functions that correspond to propagation with the speed of light.
This is consistent with the case of no discontinuities, i.e., $k_{1} = k_{2}$; then, the second term vanishes.
The integration (\ref{inttobi}) can be carried out exactly as
\begin{eqnarray}
  \dot{\phi}(t,x) &=& 
   \frac{1}{2} \left[ \delta(x-t) + \delta(x+t) \right] +  \frac{1}{2\pi (t^{2} - x^{2})} \nonumber \\
   &\times&
   \bigg[ (t+x) \Big( \sin(t-x)k_{1} - \sin(t-x)k_{2} - \sin\big[(t-x)k_{1} +Ct\big] + \sin\big[(t-x)k_{2} +Ct\big]  \Big)  \nonumber \\
      &&  + (t-x) \Big( \sin(t+x)k_{1} - \sin(t+x)k_{2} - \sin\big[(t+x)k_{1} +Ct\big] + \sin\big[(t+x)k_{2} +Ct\big]  \Big)  \bigg]. \nonumber \\
\end{eqnarray}
The second term is nonzero even when $x>t$, so that 
the wave propagates space likely.
Therefore, it is superluminal. In particular, the front velocity is infinite.

\subsection{Cuspy modulation and massive-like linear dispersion relation}
\begin{figure}[ht]  
\centering
\includegraphics[width=6cm]{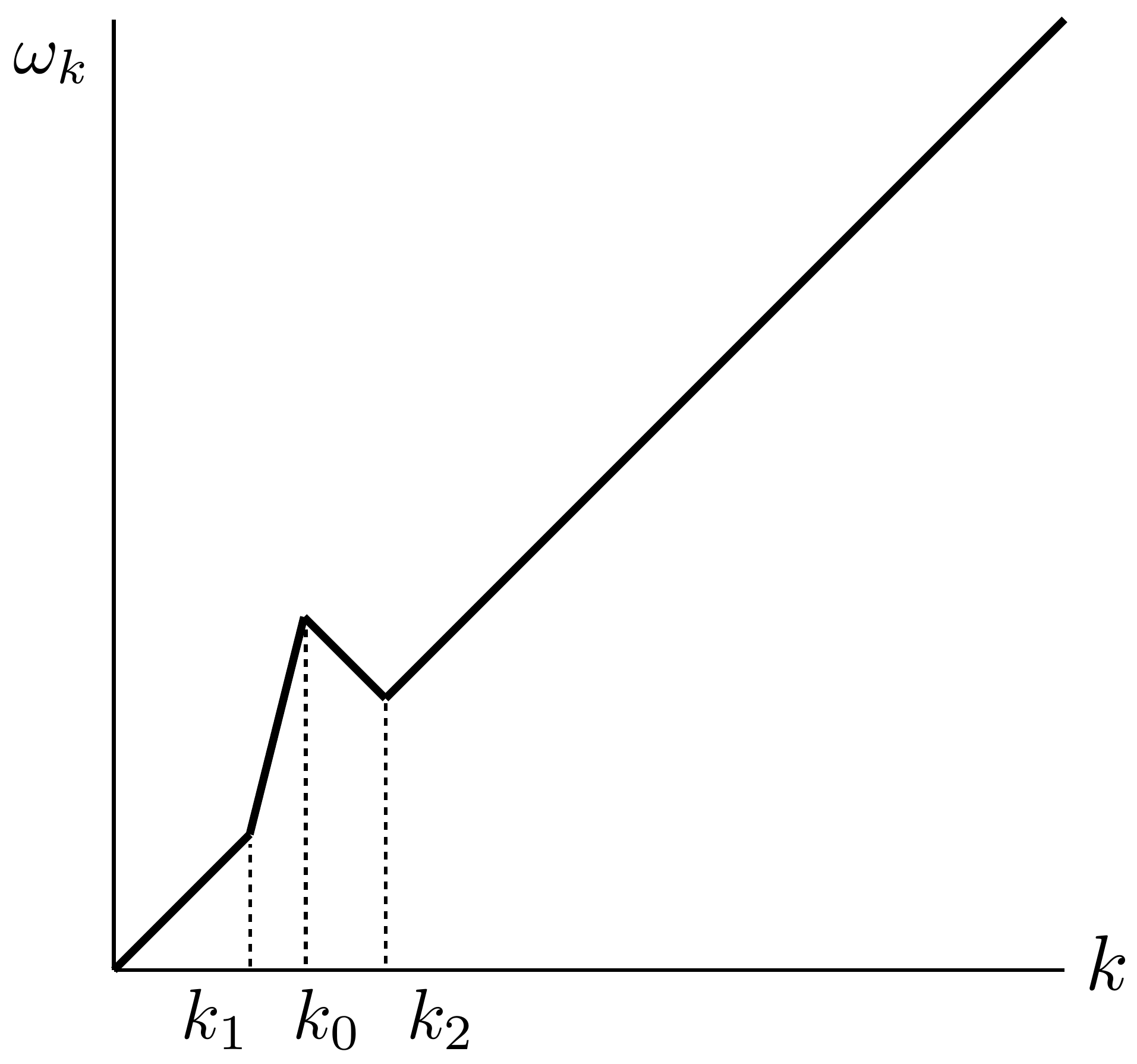}
\caption{Schematic shape of dispersion relation (\ref{dis1}).
} \label{fig_tongari}
\end{figure}

\begin{figure}[t]  
\centering
\includegraphics[width=6cm]{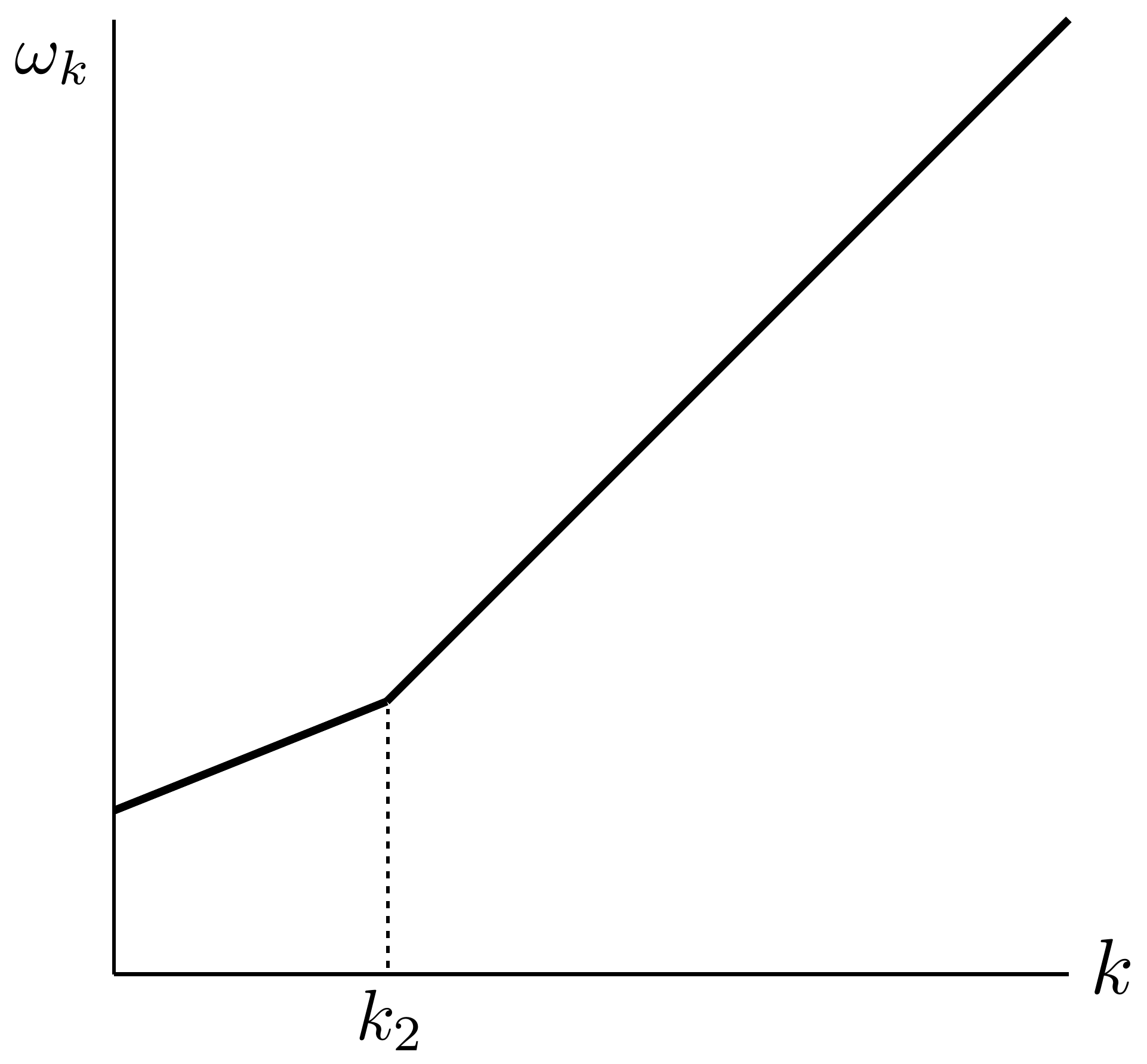}
\caption{Dispersion relation (\ref{dis1}) for $k_{0} = k_{1} = 0$.
} \label{fig_mass}
\end{figure}

Next, we consider modulation of the dispersion relation of a massless particle, as shown in Fig.\,\ref{fig_tongari}.
In Fig.\,\ref{fig_tongari}, there is a cusp with a peak at $k_{0}$ between arbitrary momenta $k_{1}$ and $k_{2}$.
More explicitly, the modulated dispersion relation for $k>0$ is given by
%
%
%
\begin{equation}
 \omega_{k} =
\begin{dcases*}
    k & for \  $ 0 < k < k_{1},~~k \ge k_{2}$ \ , \\
    ak + (1-a) k_{1}   & for \  $k_{1} \le k < k_{0}$ \ , \\
    bk + (1-b) k_{2}   & for  \   $k_{0} \le k < k_{2}$ \ ,
  \end{dcases*}  \label{dis1}
\end{equation}
where $a$ is a constant and $b = \frac{k_{2}-k_{1} - a (k_{0} - k_{1})}{k_{2}-k_{0}}$.
For dispersion relation (\ref{dis1}), Eq.\,(\ref{time}) can be evaluated as
\begin{eqnarray}
  {\dot \phi} (t,x)  &=&  \int^{\infty}_{-\infty} \frac{dk}{2\pi} e^{ikx} \cos (kt)   
             - \int^{k_{2}}_{k_{1}} \frac{dk}{\pi} \cos(kx) \cos(kt)  \nonumber \\
         &&    + \int^{k_{0}}_{k_{1}} \frac{dk}{\pi} \cos(kx) \cos\big[ (ak + (1-a)k_1)t \big]  
             + \int^{k_{2}}_{k_{0}} \frac{dk}{\pi} \cos(kx) \cos\big[(bk + (1-b)k_2) t \big]  \ .\nonumber
\end{eqnarray}
The first term gives rise to delta functions that indicate propagation with the speed of light.
The remaining terms represent the effects of the cusp in the dispersion relation and
they can be integrated exactly as
\begin{eqnarray}
  {\dot \phi} (t,x) 
    &=&   \frac{1}{2} \left[ \delta(x-t) + \delta(x+t) \right]  \nonumber \\
    && +  \frac{(t+x) \Big\{ \sin\left[ k_{1}(t-x) \right] - \sin\left[ k_{2}(t-x) \right] \Big\}
                +(x\to -x)}
               {2\pi (t^{2} - x^{2})}  \nonumber \\
    && +  \frac{(at+x) \Big\{ \sin \left[ k_{0}(at-x) + k_{1}(1-a)t \right] - \sin\left[ k_{1}(t-x) \right] \Big\}
               + (x\to -x) }
               {2\pi (a^{2}t^{2} - x^{2})}  \nonumber \\
    && -  \frac{(bt+x) \Big\{  \sin\left[ k_{0}(bt-x) + k_{2}(1-b)t \right] - \sin\left[ k_{2}(t-x) \right] \Big\}
               + (x\to -x) }
               {2\pi (b^{2}t^{2} - x^{2})}       \   .    \nonumber \\
        \label{inte1} 
\end{eqnarray}
From Eq.\,(\ref{inte1}), we find that $\dot{\phi}$ is nonzero (when $a \neq 1$) even in spacelike region $x>t$.
Therefore, the front velocity is infinite and the propagation is superluminal.
We mention that the above illustration includes a special case ($k_0=k_1=0$) where
the dispersion relation represented by Fig.\,\ref{fig_mass} resembles the one for the massive particle 
in the sense that $\omega$ near $k=0$ approaches a positive constant.
The example in this subsection also explicitly shows that IR physics can be relevant 
to the discussion of the propagation speed.

\subsection{Bumpy modulation}
\begin{figure}[t]  
\centering
\includegraphics[width=6cm]{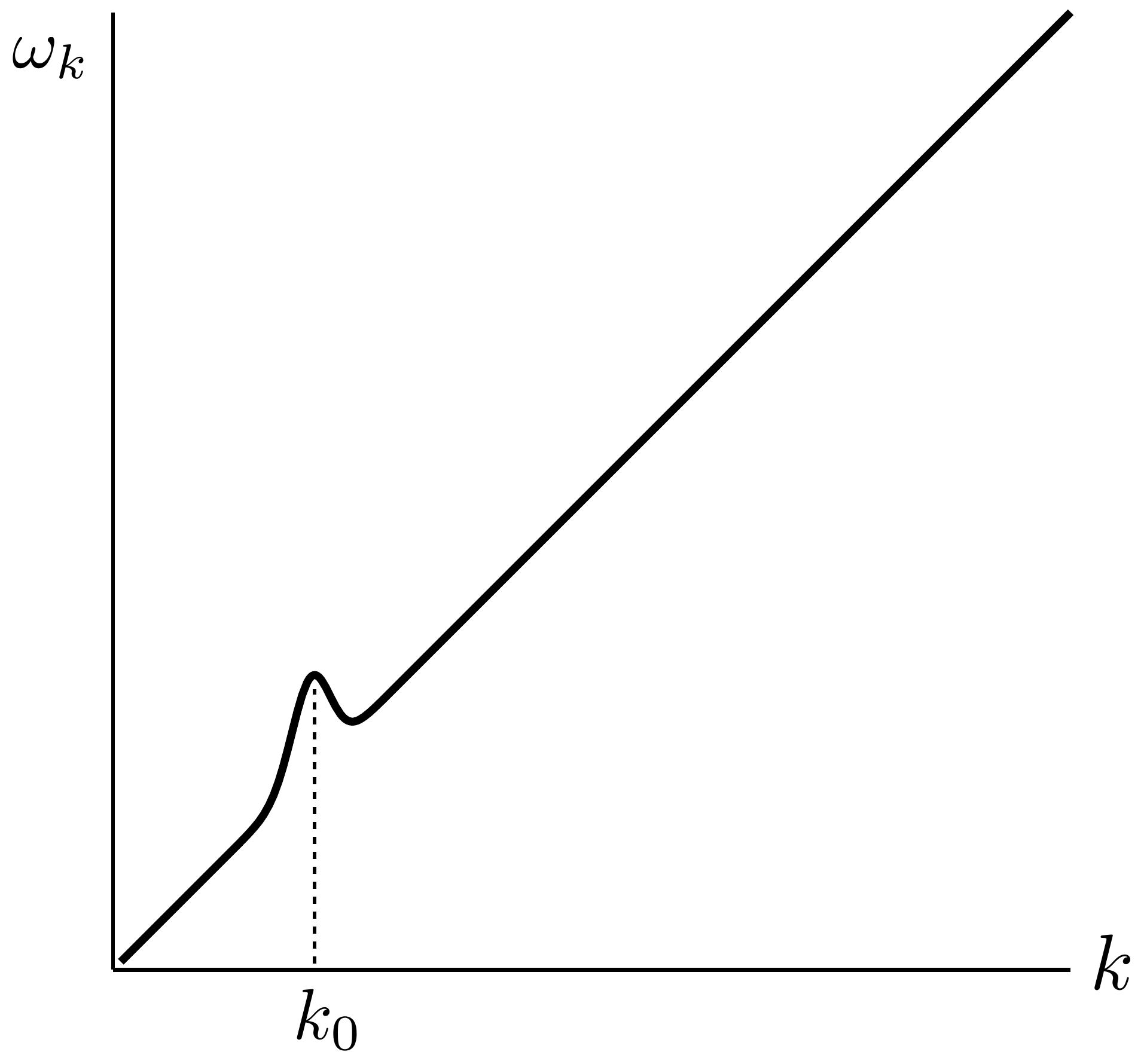}
\caption{Dispersion relation (\ref{dis}) with modulation (\ref{modu}).
} \label{fig_gau}
\end{figure}

Finally, we consider a more general modulation of the standard dispersion relation 
of a massless particle:
\begin{equation}
  \omega_{k}^{2} = k^{2} + \delta\tilde{\omega}^{2}(k) .
\end{equation}
Assuming that the modulation is small for arbitrary $k$, i.e., $\delta\tilde{\omega}(k) \ll k$, one can obtain 
\begin{equation}
  \omega_{k} \simeq k + \delta\omega(k) ,  \label{dis}
\end{equation}
for $k>0$.
We have defined $\delta\omega(k) \equiv \delta\tilde{\omega}^{2}(k) /2 k \ll k$ and
postulated parity conservation, so that $\delta\omega(-k) = \delta\omega(k)$.
In this convention, $\omega_k=-k+\delta \omega (-k)$ for $k<0$.
For the above dispersion relation (\ref{dis}),
Eq.\,(\ref{time}) can be approximated as
\begin{eqnarray}
  {\dot \phi} (t,x) 
                    &=&    \int^{\infty}_{-\infty} \frac{dk}{2\pi} e^{ikx}\cos \left( (k + \delta\omega(k))t \right) \nonumber \\
               &\simeq&  \int^{\infty}_{-\infty} \frac{dk}{2\pi} e^{ikx} \cos (kt)  
                        - \int^{\infty}_{0} \frac{dk}{\pi} \cos(kx) \sin (kt) \delta\omega(k)t  .  \label{intgau}
\end{eqnarray}
Now as an illustrative example, we consider 
a Gaussian-like function for the modulation $\delta\omega(k)$:
\begin{equation}
  \delta\omega(k) = A \exp\left( - \frac{(k-k_{0})^{2}}{2 \sigma^{2}} \right) ,   \label{modu}
\end{equation}
where $A$ is a constant.
$k_{0}$ and $\sigma$ specify a peak position and width of the distribution in momentum space, respectively.
For technical convenience, we assume $k_0 \gg \sigma$.
An example of the dispersion relation is depicted in Fig.\,\ref{fig_gau}.
%
Note that although $A$ is taken to be positive in Fig.\,\ref{fig_gau}, the signature does not change our conclusion.
The first term in Eq.\,(\ref{intgau}) gives rise to the delta functions 
that indicate propagation with the speed of light.
For the Gaussian-like function (\ref{modu}), the integration range of the second term in Eq.\,(\ref{intgau}) can be 
extended from $-\infty$ to $\infty$ approximately.
Then the integration (\ref{intgau}) can be evaluated as
\begin{eqnarray}
  {\dot \phi} (t,x) &=& \frac{1}{2} \left[ \delta(x-t) + \delta(x+t) \right]  \nonumber \\
                    &&  -\frac{\sigma A t}{\sqrt{2\pi}} 
                       \left( e^{-\sigma^{2}(t+x)^{2}/2} \sin\big[(t+x)k_{0}\big]  
                     + e^{-\sigma^{2}(t-x)^{2}/2} \sin\big[(t-x)k_{0}\big]   \right)  .         
\end{eqnarray}
We see that $\dot{\phi}$ is nonzero even in the spacelike region $x>t$, and thus, it 
means superluminal.

From the above examples A, B, and C, we find that front velocities can be superluminal even if a dispersion relation is modulated only in the IR regime, namely,
phase velocities are luminal in the UV limit.
In fact, this conclusion is quite general since the second term in Eq.\,(\ref{intgau}) indicates that
special cancellation of the effects of IR modification in a dispersion relation
is needed to avoid superluminal propagation.

\section{Conclusion}  
Requiring non-superluminal propagation is a key criterion in constructing a consistent theory.
Then it is useful to investigate front velocities that may be related to phase velocities by Eq.\,(\ref{rela}) 
to quantify the propagation speed.
Eq.\,(\ref{rela}) implies a remarkable intuition that the propagation speed of a particle is 
only determined by the UV limit of a dispersion relation, and thus, IR physics is irrelevant to it.
However, to our best knowledge, Eq.\,(\ref{rela}) was proved only for the dispersion relation given by 
Eq.\,(\ref{pow}).
In this paper, we investigated whether Eq.\,(\ref{rela}) is true for not only the dispersion relation 
of the form given by Eq.~(\ref{pow}) but also other functional forms.

In Sec.\,\ref{UV}, we showed that Eq.\,(\ref{rela}) is satisfied even for $\alpha = 4,6,8,\dots$.
This proof extends the applicable range of Eq.\,(\ref{rela}) to dispersion relations that have higher spatial
derivatives.
Therefore, the result would be beneficial to constructing low energy effective theories from the viewpoint of UV completion
where Eq.\,(\ref{rela}) plays an important role~\cite{Shore:2007um,deRham:2019ctd}.

In Sec.\,\ref{IR}, we demonstrated several examples that violate Eq.\,(\ref{rela}) by evaluating 
Green functions.
They are counterexamples to Eq.\,(\ref{rela}) and explicitly show that IR physics can be relevant to the 
discussion of propagation speed.
Our finding conveys the important caution that not only UV physics but also IR physics can be significant
when one considers a consistent theory in light of causality.

\begin{acknowledgments}
A.\,I.\ would like to thank Chong-Sun Chu, Chun-Hei Leung, Paolo Gondolo, Shinichi Hirano, and Masahide Yamaguchi for helpful discussions.
A.\,I.\ was supported by World Premier International Research Center Initiative (WPI), MEXT, Japan, and JSPS KAKENHI Grant Number JP21J00162, JP22K14034.
T.\,S.\ is supported by MEXT KAKENHI Grant Number 17H06359, JP21H05453,  
and JSPS KAKENHI Grant Number JP19K03864.
\end{acknowledgments}

\bibliographystyle{unsrt}
\bibliography{ref}

\end{document}